# DETECTING INFORMED ACTIVITIES IN EUROPEAN-STYLE OPTION TRADINGS[1]

*Lyudmila A. Glik [2], Oleg L. Kritski [3],*
*Tomsk Polytechnic University, Russia*

**Abstract**

We propose a mathematical procedure for finding informed trader activities in European-style options and their underlying asset. The regression model (9) with moving average component was written. Being added to it ARMA-process for log-price differences of underlying asset, the generalized model is written as Vector ARMA, stable at $|\rho|<1$. We also constructed an informed trader activity presence criterion. Using TAIFEX option prices we investigate whether such activity was at the market. We found that there is no significant influence for pricing process made by major market players.

## 1. Introduction

Identifying of informed traders, defined here as large institutional investors (such as pension or hedge funds, market makers, dealers and etc.), is an actual and complex problem. They play an important role in maintaining the operating performance of the stock market by providing it with a high liquidity, by keeping up some level of prices and trading volumes of financial instruments and by smoothing a significant price volatility. Besides the special functions, they use privileges which are not accessible to individual investors and speculators (Scholtus et al. (2014); Lepone and Yang (2013); Dey and Radhakrishna (2013)): they may obtain information about important macroeconomic indicators faster, they may also possess nonpublic information about firms before it is officially published in the quarterly or annual financial reports. For example, Scholtus et al. (2014) determine the statistical significance of return changes and show an increase in trading activity before the official announcement time of such important economic indicators as Chicago PMI and UMich Sentiment. It appears that subscribers receive information 3 and 5 minutes earlier than others. And, at the time of the announcement published data is already included in the price and significant spikes disappear.

---

[1] This work was financially supported by Russian Scientific Fund.
[2] Tomsk Polytechnic University, 30 Lenin Ave., Tomsk, Tomsk reg., Russia, 634050.
[3] Corresponding author. Tomsk Polytechnic University, 30 Lenin Ave., Tomsk, Tomsk reg., Russia, 634050. Tel:+73822 418913. Fax: +73822 529658.
E-mail address: olegkol@tpu.ru

The quantity of papers devoted to the detection of insider and informed traders in option and futures markets is not so large. Thus, Popescu and Kumar (2013) verify the put-call parity for American options with two different strikes for different market agents and compute the probability of informed trading (PIN). And, Hu (2013) found an order imbalance in trades of the underlying asset and its three options, i.e. in-the-money (ITM), at-the-money (ATM) и out-of-the-money (OTM) options, that enables to compute the probability of informed traders. Muravyev et al. (2013) combine those two approaches.

The possibility of addressing futures prices for detection of informed traders was first noted in Yi-Tsung (2013), where it is shown that the enhancement of the trading activity in the futures market anticipates an increased activity in the spot-market, but in most cases the direction of trade (to sell or to buy) remains unclear, even if using the non-public information.

This work develops the multi-period mixed model of monitoring insider trading activities on a stock market considered in Park and Lee (2010). We generalized it to option trading case and found the dependence between differences of function values computed at some deltas for European-style options, and underlying asset price returns. Equations obtained were written as Vector ARMA-process. The stationarity of this one was proved. At last we formulated a detection criterion of informed trading, whose effectiveness was investigated on historical data.

This study consists of the following. In Section 2 we construct a mixed multi-period model of informed trading in terms of option deltas and logarithms of stock prices series. We write model as Vector ARMA-process and investigate its characteristics. In Section 3 we analyze the effectiveness of detection criterion through a simulation and apply it to TAIFEX European options and stock index as underlying asset. Finally, Section 4 presents our conclusions.

## 2. The model

Let's assume that number of all players on the market, trading an underlying asset and their put and call options, are divided into informed traders and ordinary "noise" traders. Let a macroeconomic announcement affecting the price, be publicly known at a future time $T$, while the informed trader has obtained the data at time $t<T$. We assume that he decides to buy (sell) the underlying asset or options in equal installments at regular intervals, i.e. at time $t$, $(t+1)$, ..., $T$. Then the effect of changes in the underlying asset price could be defined as

$$X_t = v_t + u_t,$$

where $u_t \sim N(0, \sigma_u^2)$ – addition to the price, offered by uninformed traders, $v_t$ – surcharge to the price that an informed trader is willing to pay.

Let $v_t$ obey the relation

$$v_t = \beta \Psi_t,$$

where $\beta$ – coefficient of proportionality, $\Psi_t = \ln \theta_t$ – nonlinear asset price dynamics, $\theta_t$ – value of trading transaction of informed trader.

We suppose that $\Psi_t$ satisfies AR(1) model, which is explained by the desire of the informed trader to hide his activity and, for instance, to reduce his contribution at low levels of market activity:

$$\Psi_t = \overline{\Psi} + \rho \Psi_{t-1} + z_t, \qquad (1)$$

where $\overline{\Psi}$ – average logarithmic value of trading transaction price per time unit, $z_t \sim N(0, \sigma_z^2)$ – price noise.

Let $S_t$ be a quote of the underlying asset at time $t$. Since a trader buys a large quantity of the underlying asset, we assume that log-price $\ln S_t$ will vary proportionally to the change of prices:

$$\ln S_t = \ln S_{t-1} + \lambda X_t, \qquad (2)$$

where $\lambda = \dfrac{\mathrm{cov}((\Psi_t, X_t)|v_{t-1})}{D(X_t|v_{t-1})} = \dfrac{\beta \sigma_z^2}{\beta \sigma_z^2 + \sigma_u^2}$ – market depth variation coefficient, if noise processes $z_t, u_t$ are independent of one another.

Denote $R_t = \ln S_t$, so, we can rewrite (2) in more convenient way as

$$\Delta R_t = \lambda X_t, \qquad (3)$$

that describes log-return price dynamics for underlying asset. The expression (3) coincides with model of asset price differences investigated in Park and Lee (2010). So, we can state our main results of this paper for $\Delta R_t$.

**Theorem** 1. *In case an informed trader implements a mixed strategy* (1), (3) *with* $\lambda = \dfrac{4\beta \sigma_z^2}{4\beta \sigma_z^2 + \sigma_u^2}$ *as the information exposure strategy in a multi-period model, the logarithmic price returns $\Delta R_t$ follows the ARMA(1,1)-process below*

$$\Delta R_t = \gamma + \rho \Delta R_{t-1} + \varepsilon_t + \delta \varepsilon_{t-1}, \qquad (4)$$

*where* $\varepsilon_t \sim N(0, \sigma_\varepsilon^2)$ – *noise*, $\delta = \left[\sigma_u^2(1+\rho^2) + 2\beta^2 \sigma_z^2 - (1+\rho)\sigma_u \sqrt{4\beta^2 \sigma_z^2 + \sigma_u^2(1-\rho)^2}\right]\left(2\rho\sigma_u^2 - 2\beta^2 \sigma_z^2\right)^{-1}$,

$\sigma_\varepsilon^2 = \lambda^2 \beta^2 \sigma_z^2 (1+\rho^2)(\rho\delta^2 + \rho^2\delta + \rho + \delta)^{-1}$, $\gamma = \lambda\beta(1-\rho)\dfrac{\ln S_T - \ln S_0}{T}$.

The proof is listed in the ***Appendix A***.

Use the nonlinear price dynamics written as equality (3) to discover an informed trading activity in European-style option trades.

It is well-known that fair price of European-style option with expiry date $T$, implied volatility $\sigma$, underlying asset price $S_t$, expiration price $E$ and risk-free interest rate $r$ follows the famous Black-Scholes formula (Hull (2003)) at any time $t$:

$$V_t = S_t \Phi(d_1) - E e^{-r(T-t)} \Phi(d_2),$$

where $d_1 = \dfrac{\ln S_t - \ln E + (r + \sigma^2/2)(T-t)}{\sqrt{(T-t)}\sigma}$, $d_2 = d_1 - \sigma\sqrt{T-t}$, $\Phi(x) = \int_{-\infty}^{x} \dfrac{1}{\sqrt{2\pi}} e^{-t^2/2} dt$, $0 \leq t \leq T$.

Because variables $V_t$ and $d_i$, $i=1,2$, depend on $S_t$ in a non-linear way, it is impossible to write model for differences $\Delta V_t$ like it was made in expressions (3) and (4). So, we will use classical hedging coefficients $\Delta_t = \dfrac{\partial V_t}{\partial S_t}$ to obtain all equations of our model.

**2.1. European style call option**

It is well-known that delta for European-style call option is equal to

$$\Delta_t = \Phi(d_{1,t}),$$

where $d_{1,t} = \dfrac{\ln S_t - \ln E + (r + \sigma^2/2)(T-t)}{\sqrt{(T-t)}\sigma}$. Therefore if we take a reversed to normal p.d.f. function $\Phi^{-1}(y)$ from all parts of expression above, we get

$$d_{1,t} = \Phi^{-1}(\Delta_t),$$

$$d_{1,t-1} = \Phi^{-1}(\Delta_{t-1}),$$

i.e., denoting $Q_t = \Phi^{-1}(\Delta_t)$, we obtain

$$\Delta d_{1,t} = \Delta Q_t. \qquad (5)$$

We take advantage of Taylor series expansion of the function $\sqrt{1+\tau}$ near the point $\tau_0 = \tau - 1$:

$$\sqrt{1+\tau} = \sqrt{\tau} + \dfrac{1}{2\sqrt{\tau}} + O\left(\dfrac{1}{\sqrt{\tau^3}}\right).$$

Then

$$\sqrt{1+T-t} = \sqrt{T-t} + O\left(\dfrac{1}{\sqrt{T-t}}\right). \qquad (6)$$

Applying (6) to computing an approximation of $d_{1,t-1}$, we find, that

$$d_{1,t-1} = \dfrac{R_{t-1} - \ln E + (r + \sigma^2/2)(T-t+1) + O((T-t)^{-1})}{\sqrt{(T-t)}\sigma},$$

i.e.
$$\Delta d_{1,t} = \frac{\Delta R_t - (r + \sigma^2/2) + O((T-t)^{-1})}{\sqrt{(T-t)}\sigma}. \tag{7}$$

So, the final form of expression (5) is as follows

$$\Delta Q_t = \frac{\Delta R_t - (r + \sigma^2/2) + O((T-t)^{-1})}{\sigma\sqrt{(T-t)}}. \tag{8}$$

Substitute (8) to (4) for taking into account the underlying asset logarithmic return dynamics:

$$\Delta Q_t = \frac{\gamma - (r + \sigma^2/2) + O((T-t)^{-1})}{\sigma\sqrt{(T-t)}} + \frac{\rho}{\sigma\sqrt{(T-t)}}\Delta R_{t-1} + \frac{1}{\sigma\sqrt{(T-t)}}\varepsilon_t + \frac{\delta}{\sigma\sqrt{(T-t)}}\varepsilon_{t-1}. \tag{9}$$

We note that expression (9) is the desired regression model with moving average, that takes into account quantile differences of normal p.d.f. computed in points $\Delta_t$ and $\Delta_{t-1}$. The model also depends implicitly from call option price differences. This dependence allows us to find the informed traders activity. To do so, we write the system of equations (4) and (9) as Vector ARMA-process and find a condition of its stationarity. The Theorem 2 holds:

**Theorem 2.** *System of equalities* (4), (9) *can be written in the form of the Vector ARMA:*

$$X_t = A_t + X_{t-1}B + \Gamma_t \varepsilon_{t-1} + H_t \varepsilon_t, \tag{10}$$

where $X_t = (\Delta Q_t; \Delta R_t)$, $A_t = \left(\frac{\gamma - (r + \sigma^2/2) + O((T-t)^{-1})}{\sigma\sqrt{(T-t)}}; \gamma\right)$, $\Gamma_t = \left(\frac{\delta}{\sigma\sqrt{T-t}}; \delta\right)$,

$H_t = \left(\frac{1}{\sigma\sqrt{T-t}}; 1\right)$ *are vectors,* $B = \begin{pmatrix} 0 & 0 \\ \frac{\rho}{\sigma\sqrt{T-t}} & \rho \end{pmatrix}$ – *square matrix* $2 \times 2$.

The proof of Theorem 2 is obvious.

We note that representing the process in the form of (6) allows us to determine the conditions of its stationarity (see, e.g., Wei (2006) for getting conditions and estimation of the stationarity of Vector ARMA processes). From the results we can formulate Theorem 3:

**Theorem 3.** *Process $X_t$ in expression (10) is stationary, if $|\rho| < 1$.*

The proof of Theorem 3 is listed in *Appendix A*.

Let consider a European-style put option.

## 2.2. European style put option

It is well-known that delta for European-style put option is equal to

$$\Delta_t = \Phi(d_{1,t}) - 1.$$

Therefore if we denote $\tilde{Q}_t = \Phi^{-1}(\Delta_t + 1)$ and take a reversed to normal p.d.f. function $\Phi^{-1}(y)$ from all parts of expression above, we get

$$\Delta d_{1,t} = \Delta \tilde{Q}_t.$$

Further, for put option the expressions (7) and (4) hold, so, the last equality can be rewritten as follows:

$$\Delta \tilde{Q}_t = \frac{\gamma - (r + \sigma^2/2) + O((T-t)^{-1})}{\sqrt{(T-t)}\sigma} + \frac{\rho}{\sqrt{(T-t)}\sigma}\Delta R_{t-1} + \frac{1}{\sqrt{(T-t)}\sigma}\varepsilon_t + \frac{\delta}{\sqrt{(T-t)}\sigma}\varepsilon_{t-1},$$

and it looks very similar to (9). And so, of course, Theorems 2 and 3 will also be true for European-style put option with $X_t = (\Delta \tilde{Q}_t; \Delta R_t)$.

## 2.3. Criterion of presence of informed trading activity

If process $X_t$ is stationary, the expression (9) allows us to formulate criteria, helping to find informed trader activities on stock and option markets. For instance, if their influence is greater than price return noise then coefficient before logarithmic differences $\Delta R_{t-1}$ in (9) is much greater than coefficient before $\varepsilon_{t-1}$. Hence, we consider that informed tradings are found if the absolute value of $\frac{\rho}{\sigma\sqrt{(T-t)}}$ is greater than $\frac{\delta}{\sigma\sqrt{(T-t)}}$ at any time $t$, but their signs should be opposite, i.e. 1) if $\rho<0$, then $0<\delta<-\rho$; 2) if $\rho>0$, then $-1<\delta<-\rho$. This is the **criterion of informed trading** to be formulated.

For empirical high frequency values of stock and option prices we should use some statistical estimates of $\hat{\rho}$ and $\hat{\delta}$. Let $R_t, t = 0,1,..,(T-1)$, be a data set that is available for analysis. Let $m<<T$ be a time window length, which allows us to calculate the initial estimates of coefficients $\hat{\gamma}_1 = \left( \hat{\lambda}_1 \beta (1-\hat{\rho}_1) \frac{\ln S_m - \ln S_0}{m} \right)$, $\hat{\rho}_1$ and $\hat{\delta}_1$ with values $R_0, R_1, R_2,..., R_m$ in model equations (4) and (9). Moving the time window to the right per unit until we reach time $(T-1)$, by known $R_s, R_{1+s},..., R_{m+s}$ we estimate $\hat{\gamma}_s$, $\hat{\rho}_s$ and $\hat{\delta}_s$, $s = 0,1,..,(T-m-1)$.

Furthermore, we use empirical values of the coefficients found to formulate a decision rule convenient to practical computations. Accordingly, we write the criterion.

**Decisive criterion**: subject to non-zero price movements $\hat{\gamma}_k \neq 0$, $k = 0,1,..,(T-m-1)$ and satisfy the inequality $|\hat{\rho}_k| < 1$, we assume that the informed transaction is detected if one of the following inequalities holds:

a) $\sum_k \hat{\rho}_k \sum_k \hat{\delta}_k < 0$, $\sum_k \hat{\rho}_k < 0$, $\left|\sum_k \hat{\rho}_k\right| > \left|\sum_k \hat{\delta}_k\right|$; б) $\sum_k \hat{\rho}_k \sum_k \hat{\delta}_k < 0$, $\sum_k \hat{\rho}_k > 0$, $\left|\sum_k \hat{\rho}_k\right| < \left|\sum_k \hat{\delta}_k\right|$.

## 3. Numerical simulation results

The most of the world financial markets have no European style options for trading, but American ones only. So, for making numerical investigation of proposed model (10) we use historical data of Taiwan Futures Exchange TAIFEX, free for downloading[4]. Moreover, it provides for free use delta values for all kinds of derivatives traded, that is very convenient for our estimations of model coefficients in (4) and (9).

We take week options call and put (ticker: TXO for all of them) on index TSEC (ticker TPE:TAIEX), the contracts expire at 26/02/14. Strike values were chosen in such a way as to implied volatility (*IV*) possesses the minimum value while a quantity of trade deals reaches a maximum. We state such parameters for call week option: $T=7/360$, $t=1/360,…, 6/360$, $E=8650$, $IV=20,87\%$, $\Delta_1=0,19$; $\Delta_2=0,38$; $\Delta_5=0,29$. For put week option they are as follows: $T=7/360$, $t=1/360,…, 6/360$, $E=8500$, $IV=20,87\%$, $\Delta_1=-0,49$; $\Delta_2=-0,27$; $\Delta_5=-0,35$.

Model coefficients in (10) were estimated at time $t=5/360$. We found out for call option that $\frac{\hat{\rho}}{\sigma\sqrt{(T-t)}}=-5,54$, $\frac{\hat{\delta}}{\sigma\sqrt{(T-t)}}=-26,7$. For put option they are -1,34 and -37,9 correspondingly. So, for autoregression process (4) the coefficients to be estimated are $\hat{\rho}=-0,21$ and $\hat{\delta}=-1,0$, i.e. in accordance with Theorem 3 proved the model (10) is stationary for call and put options. But decisive criterion declines the presence of informed trader activities because $\hat{\rho}$ and $\hat{\delta}$ have like signs.

## 4. Conclusion

We suggested the mathematical procedure of finding informed trader activities in stock and European option markets. Computations made for derivatives and their underlying assets of Taiwan Futures Exchange showed that principles of fair and honest trading are fulfilled. And no suspect activity was found.

*APPENDIX A*. **Proofs**

*Proof of Theorem 1.* After double substitution of equation (1) into (3) with different *t* we obtain:

$$\Delta R_t = \gamma + \rho\lambda\beta\Psi_{t-1} + \lambda\beta z_t + \lambda\beta z_{t-1} + \lambda u_t, \qquad (A.1)$$

$$\Delta R_{t+1} = \gamma(1+\rho) + \rho^2\lambda\beta\Psi_{t-1} + \lambda\beta(1+\rho)z_t + \lambda\beta\rho z_{t-1} + \lambda\beta z_{t+1} + \lambda u_{t+1}. \qquad (A.2)$$

We modify (A.2) as

$$\Delta R_{t+1} = \gamma + \rho\Delta R_t + \lambda\beta z_{t+1} + \lambda\beta z_t + \lambda u_{t+1} - \lambda\rho u_t. \qquad (A.3)$$

We compute autocovariance of some auxiliar expression

---
[4] http://www.taifex.com.tw/eng/eng3/eng3_4.asp

$$\Delta R_t = \gamma + \rho \Delta R_{t-1} + \varepsilon_t + \delta \varepsilon_{t-1},$$

at times $t=0$ and $t=1$ and obtain its values, denoted as $V_0$ and $V_1$:

$$V_0 = \sigma_\varepsilon^2 (1+\delta^2 + 2\rho\delta)(1-\rho^2)^{-1}, \tag{A.4}$$

$$V_1 = \sigma_\varepsilon^2 (\rho + \rho\delta^2 + \rho^2\delta + \delta)(1-\rho^2)^{-1}. \tag{A.5}$$

If we compute similar autocovariances $V_0$ and $V_1$ of expression (A.3) and equate them to (A.4), (A.5), we can find unknown $\delta$ and $\sigma_\varepsilon^2$ solving to the following system

$$\sigma_\varepsilon^2 (1+\delta^2 + 2\rho\delta)(1-\rho^2)^{-1} = \lambda^2 \beta^2 \sigma_z^2 (1-\rho)^{-1} + \lambda^2 \sigma_u^2,$$

$$\sigma_\varepsilon^2 (\rho + \rho\delta^2 + \rho^2\delta + \delta)(1-\rho^2)^{-1} = (1+\rho)\lambda^2 \beta^2 \sigma_z^2 (1-\rho)^{-1}.$$

It can be shown easily that

$$\delta = \left[\sigma_u^2(1+\rho^2) + 2\beta^2\sigma_z^2 - (1+\rho)\sigma_u\sqrt{4\beta^2\sigma_z^2 + \sigma_u^2(1-\rho)^2}\right]\left(2\rho\sigma_u^2 - 2\beta^2\sigma_z^2\right)^{-1},$$

$$\sigma_\varepsilon^2 = \lambda^2 \beta^2 \sigma_z^2 (1+\rho^2)(\rho\delta^2 + \rho^2\delta + \rho + \delta)^{-1}.$$

These coefficients, obviously, define our auxiliar expression above in the form (4) we need.

*Proof of Theorem 3.* According to the general theory of stationary of Vector ARMA processes (Wei (2006)), $X_t$ is stationary if and only if all eigenvalues of B lie inside the unit circle in the complex plane. Clearly, matrix B has two real eigenvalues: zero and $\rho$. Since $|\rho|<1$ by condition of the theorem, then $X_t$ is stationary.

# References

Dey, M.K., Radhakrishna, B., 2013. Informed trading, institutional trading, and spread. Journal of Economics and Finance, 1-20.

Hu, J., 2014. Does option trading convey stock price information? Journal of Financial Economics. 111 (3), 625–645.

Hull, J., 2003. Options, Futures, and Other Derivatives. New Jersey: Prentice-Hall, Saddle River, 5$^{th}$ edition, 755 p.

Lepone, A., Yang, J.Y., 2013. Informational role of market makers: The case of exchange traded CFDs. Journal of Empirical Finance 23, 84-92.

Muravyev, D., Pearson, N.D., Broussard, J.P., 2013. Is there price discovery in equity options? Journal of Financial Economics 107(2), 259-283.

Park, Y.S., Lee, J., 2010. Detecting insider trading: The theory and validation in Korea Exchange. Journal of Banking & Finance 34 (9), 2110-2120.

Popescu, M., Kumar, R., 2013. The implied intra-day probability of informed trading. Review of Quantitative Finance and Accounting, 1-15.

Scholtus, M., Dick, van Dijk, Frijns, B., 2014. Speed, algorithmic trading, and market quality around macroeconomic news announcements. Journal of Banking & Finance 38, 89-105.

William, W.S., Wei, 2006. Time Series Analysis: Univariate and Multivariate Methods. Pearson Education Inc, 2$^{nd}$ edition, 614 p.

Yi-Tsung, Lee, Wei-Shao, Wu, Yang, Y.H., 2013. Informed Futures Trading and Price Discovery: Evidence from Taiwan Futures and Stock Markets. Asia-Pacific Financial Markets 20 (3), 219-242.